\documentclass[prb,preprint,final]{revtex4}
\usepackage{amsmath}
\usepackage{amsfonts}
\usepackage{graphicx}

\begin{document}

\title{Optimal design of nanoplasmonic materials using genetic algorithms as
a multi-parameter optimization tool}
\author{Joseph Yelk,$^1$ Maxim Sukharev,$^2$ and Tamar Seideman$^{1,2}$ }
\affiliation{$^1$Department of Physics and Astronomy\\
$^2$Department of Chemistry, Northwestern University, 2145 Sheridan Road,
Evanston, IL. 60208-3113 USA\\
}

\begin{abstract}
An optimal control approach based on multiple parameter genetic algorithms
is applied to the design of plasmonic nanoconstructs with pre-determined
optical properties and functionalities. We first develop nanoscale metallic
lenses that focus an incident plane wave onto a pre-specified, spatially
confined spot. Our results illustrate the role of symmetry breaking and the
mechanism of energy flow through wires and cavities. Next we design a
periodic array of silver particles to modify the polarization of an
incident, linearly-polarized plane wave in a desired fashion while
localizing the light in space. The results provide insight into the
structural features that determine the birefringence properties of metal
nanoparticles and their arrays. Of the variety of potential applications
that may be envisioned, we note the design of nanoscale light sources with
controllable coherence and polarization properties that could serve for
coherent control of molecular, electronic, or electromechanical dynamics in
the nanoscale.
\end{abstract}

\maketitle

\section{Introduction}

\label{introduction}

The rapid serge of interest in nanoplasmonics\cite{reviews} in recent years
owes largely to a broad variety of already demonstrated and projected
applications in fields ranging from information processing\cite{MaierBook07}
and integrated optics,\cite{AtwaterReview05} through single molecule
spectroscopies\cite{VanDuyneScience04} and detection,\cite{HlaReview05} to
biosensing and medical diagnostics.\cite{VanDuyneReview04} Underlying these
applications is the ability of plasmonic materials composed of nanoparticles
(NPs) to strongly enhance and spatially focus an incident electromagnetic
(EM) field at resonant frequencies and to coherently transport light in
space.\cite{BarnesReview07} An external EM field tuned to a resonance
frequency, excites collective oscillations of conductive electrons in the
skin-layer of particles. Such coherent oscillations, termed localized
surface plasmons, are confined within a small volume and hence give rise to
substantial enhancement of the total field on the surface of the NP. In an
array of metal NPs, the surface plasmon resonance enhances the interaction
between adjacent particles and may thus give rise to collective response to
external light, where near-field coupling between closely spaced particles
sets up coupled plasmon modes.

It has been realized that the optical properties of metal NPs and their
arrays depend sensitively on their size, shape and relative arrangement as
well as on the fine details of their material composition. Together with the
availability of advanced experimental methods for fabricating nanoconstructs
with nanometer precision and regularity,\cite{VanDuyneJPCB2001} this
sensitivity calls for a numerical design tool that would be able to provide
the precise structural details of a construct that will optimize a specific
property (or simultaneously optimize several properties), and hence provide
a certain desired functionality. The potential of such a tool in
nanophotonics research and technology goes beyond the availability of a
method of determining ahead of experiment or simulation the structural
parameters required for a specific functionality and hence circumventing an
expensive and inevitably incomplete trial and error procedure. With proper
construction, such design tool provides also a route to material properties
and new insights regarding plasmon wave propagation and the interaction of
near fields with matter.

Such an approach was introduced in Refs. [%
\onlinecite{SukharevNanolett06,SukharevJCP06,SukharevJPB07}] , where the
combination of genetic algorithms (GA) with a numerical solution of the
Maxwell equations was proposed as a stable design tool for nanoplasmonics.
The approach was tested by application to simple 2D test models, leading to
interesting new insights regarding the role of coherence, the potential of
the source polarization and the time evolution of plasmon waves via
composite constructs.

Experimental techniques capable of taking a numerically designed
nanoconstruct and reproducing it in the laboratory are rapidly developing.
In addition to the (now considered "traditional", but yet surprisingly
powerful) chemical and electron beam lithography approaches, advanced
methods such as the focused ion beam and the atomic force microscope are
becoming available. It is thus relevant to test the ability of the GA-based
approach to solve realistic design problems involving a large number of
parameters. It is important also to understand the convergence properties of
the approach and its mechanism of operation. Further, perhaps the most
important outcome of the studies of Refs. [%
\onlinecite{SukharevNanolett06,
SukharevJCP06}] is the realization that the polarization and phase
properties of the incident light play a major role in nanoplasmonics, hence
the potential of elliptically polarized sources whose polarization and phase
are optimized via the GA for a specific purpose. Related is the recently
established ability of nanoconstructs to modify the polarization properties
of the incident EM field. Although 2D models can be usefully applied to
various design problems, use of polarization requires the application of the
GA in full three dimensions.

Our goal in the present work is thus multifold. First, we apply the GA to a
many parameter design problem, namely the auto-design of a lens with
predetermined properties starting with an arbitrarily-shaped slab and
subject to different constraints. Here we explore the effects of symmetry
breaking and the essential properties of the class of structures that are
able to spatially localize light. Next we explore the possibility of using
the GA to design a construct that will modify the incident polarization in a
pre-determined fashion. Here we test the applicability of the algorithm in
full three dimensions.

The next section briefly reviews the theory and its numerical implementation
and Sec. \ref{results} presents and discusses our results. We conclude this
work in Sec. \ref{conclusions} with an outlook for future research.

\section{Theory and numerical implementation}

\label{model} The scattering of EM radiation by metal nanostructures is
studied using a finite-difference time-domain (FDTD) scheme\cite{TafloveBook}
to solve the Maxwell equations,
\begin{eqnarray}
\varepsilon _{\text{eff}}\frac{\partial \vec{E}}{\partial t} &=&\nabla
\times \vec{H}-\vec{J},  \notag \\
\mu _{0}\frac{\partial \vec{H}}{\partial t} &=&-\nabla \times \vec{E},
\label{Maxwell equations} \\
\frac{\partial \vec{J}}{\partial t} &=&\alpha \vec{J}+\beta \vec{E},  \notag
\end{eqnarray}%
where $\vec{E}$ and $\vec{H}$ are the electric and magnetic components of
the EM field, $\varepsilon _{\text{eff}}$ is an effective dielectric
permittivity, $\mu _{0}$ is the magnetic permeability, and $\vec{J}$ denotes
the current density.\cite{GreyPRB03} The constants $\alpha $ and $\beta $
account for material dispersion in metallic regions of space,\cite%
{Ziolkowski95} which is described in the present work within the Drude model
with a frequency-dependent dielectric constant given by,
\begin{equation}
\varepsilon \left( \omega \right) =\varepsilon _{\infty }-\frac{\omega
_{p}^{2}}{\omega ^{2}+i\Gamma \omega }.  \label{Drude model}
\end{equation}%
In Eq. (\ref{Drude model}), $\varepsilon _{\infty }$ is the infinite
frequency limit of the dielectric permittivity of metal, $\omega _{p}$ is
the bulk plasma frequency, and $\Gamma $ is the relaxation rate. In this
work we consider silver structures with the parameters provided in Ref. [%
\onlinecite{AgDrude}]. In order to simulate open systems, perfectly matched
layer (PML) boundaries\cite{Berenger} are implemented with the exponentially
differenced equations\cite{SadikuBook} and a thickness of $16$ spatial
steps. The simulation of periodic constructs applies periodic boundary
conditions in the lateral directions and PML boundaries in the direction
perpendicular to the light propagation direction.

To conclude this section we briefly outline the implementation of the GA in
the present research program and its application to the optimal design of
nanoplasmonics. We refer the reader to Ref. [\onlinecite{HauptGAbook}] for a
general review of the GA, and provide here only the details essential for
understanding the results of the next section.

GAs form a class of search techniques inspired by the biological process of
evolution by means of natural selection, which can be used as a numerical
optimization tool. The stochastic nature of the GA makes it a numerically
stable optimizer, but is also the reason for which it requires a large
number of evaluations in order to find a global extremum. In the present
context the evaluation is computationally expensive but lends itself to
efficient parallelization. It is thus practical while nontrivial.

Within the genetic algorithm approach, we consider, in the plasmonics
context, a general nano-construct composed of metallo-dielectric NPs, the
optical properties of which depend on a set of adjustable parameters. These
include in general both field parameters (such as the laser pulse duration,
wavelength, polarization, and pulse shape) and material parameters
(including the size, shape, and relative arrangement of individual NPs). The
standard GA procedure involves the following steps:

\begin{enumerate}
\item Start by generating a set (\textquotedblleft
population\textquotedblright ) of possible solutions and evaluate the
fitness function of each member of the set, where the fitness function is a
time-averaged EM energy, for instance.

\item Select pairs of solutions (``parents'') from the current set, with the
probability of a given solution being selected made proportional to that
solution's fitness.

\item Breed the two solutions selected in (2) and produce two new solutions
(``offspring'', or ``a new generation'').

\item Repeat steps (2) and (3) until the number of new solutions produced
equals the number of individuals in the current set.

\item Use the new set to replace the old one and repeat steps (1) through
(5) until a (problem-dependent) termination criterion is satisfied.
\end{enumerate}

Each individual is encoded by one additional numerical parameter, the number
of significant digits, referred to as the number of genes.

All simulations have been performed on IBM BlueGene/L cluster at Argonne
National Laboratory.

\section{Results and Discussion}

\label{results}

We begin (Sec. \ref{lens}) by discussing the design of a nanoscale silver
lens using two dimensional GA optimizations. In Sec. \ref{cross} we
illustrate the design of a construct that modifies the polarization of an
incident field in a predetermined fashion using a 3D implementation of the
GA approach.

\subsection{An optimal nanolens}

\label{lens}

\begin{figure}[tbph]
\centering\includegraphics[width=\linewidth]{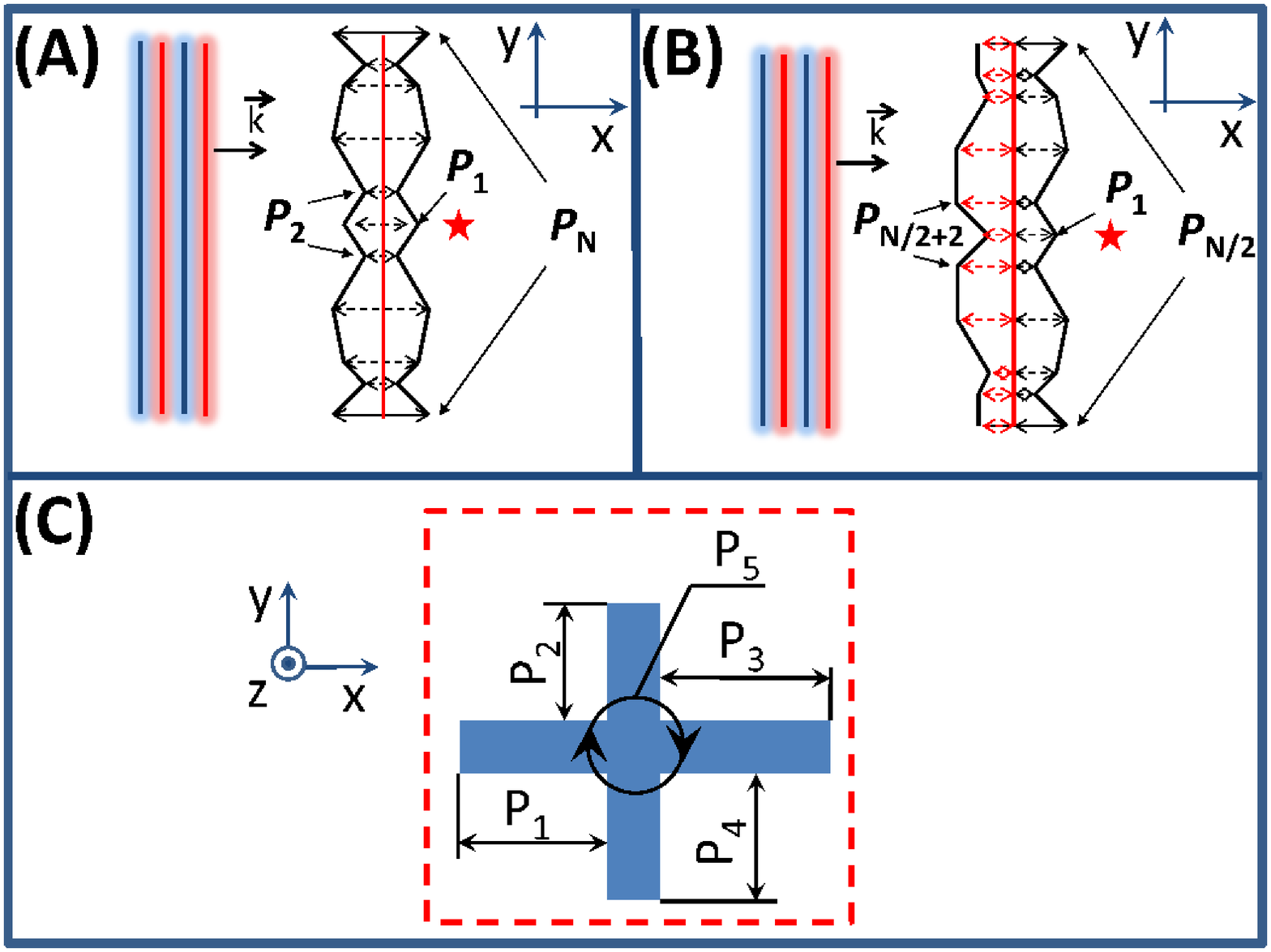}
\caption{(Color online) Schematic setup of the GA optimizations.
Panel (A) shows the setup and parameters definition, $P_{n}$, for
the design of a two-dimensional symmetric silver lens (Sec.
\protect\ref{lens}). An incident
EM plane wave polarized along the $y$-axis and propagating along the $x$%
-axis emanates from the vertical line on the left of the structure. The
fitness function is defined as the time-averaged EM energy calculated at the
point shown as a red star to the right of the lens at $58$ nm from the lens
center (normalized with respect to the incident EM energy). Panel (B) shows
a similar lens to the one shown in panel (A), where, however, symmetry with
respect to the $x=0$ plane is not imposed ($P_{N/2+1}\rightarrow P_{N-1}$
are varied independently of $P_1\rightarrow P_{N/2}$). The (arbitrarily
chosen) focus of he lens is placed to the right of the lens, $67$ nm from
the lens center, as shown by a red star. Panel (C) depicts a unit cell in
the $xy$-plane of a three-dimensional periodic structure optimized to change
the light polarization in a predetermined fashion. An Incident EM plane wave
polarized along the $x$-axis and and propagating along the $z$-axis is
generated $5$ spatial steps bellow the upper PML region. The fitness
function is defined in terms of the birefringence of the crystal discussed
in the text, as the ratio of time averaged EM energy in the induced and
incident field components, $E_{y}^{2}/E_{x}^{2}$.}
\label{fig1}
\end{figure}

Our objective in this subsection is to design an optimally shaped NP that
would focus an incident plane wave onto a pre-specified detection point.
Strong enhancement and focussing of the EM energy on the surface of the
construct is guaranteed essentially independent of the design, but our goal
here is to design a true lens, that is, to focus the light onto a specific
but arbitrary point remote from the construct. GA optimization to that end
is carried out in 2D, were attention is restricted to the
transverse-electric mode TE with two electric, $E_{x}$ and $E_{y}$, and one
magnetic, $H_{z}$, field components. We consider EM wave scattering by a
single silver NP of an arbitrary shape, as schematically depicted in Figs. %
\ref{fig1}A and \ref{fig1}B. The incident plane wave propagates from left to
right and is polarized along the vertical axis. As the fitness function in
the GA optimizations we use the time-averaged EM energy at the detection
point, normalized with respect to the incident EM energy. Since a single
evaluation of the fitness function in two dimensions is relatively fast, we
employ a different approach from that used in Refs. [%
\onlinecite{SukharevNanolett06, SukharevJCP06, SukharevJPB07}], using a
single processor to evaluate a single individual in the current generation.
The overall number of individuals in the GA optimizations is $128 $, which
results in rapid convergence, as illustrated below.

We consider two qualitatively different cases, a lens that is constrained to
be symmetric with respect to the plane perpendicular to the light
propagation vector, Fig. \ref{fig1}A, and one where the algorithm is free to
break this symmetry, Fig. \ref{fig1}B. The symmetric lens is a polygon
consisting of $42$ vertices. The positions of $11$ of the vertices serve as
parameters in the GA optimization. The $y$ (vertical) coordinate is spaced
in regular intervals of $10$ nm and the $x$ (horizontal) coordinate is a
free parameter, restricted to the $7.5$ to $22.5$ nm range (mapped to a grid
with a spatial step of $1$ nm). The positions of $11$ other vertices are
determined by reflecting the first $11$ vertices about the vertical plane
shown in red in Fig. \ref{fig1}A. The remaining $20$ vertices are determined
by reflecting the resulting $22$ vertices about the plane intersecting the
bottom two vertices (the bottom two are reflected onto themselves). The
total number of parameters in the GA optimizations is $12$, including the $%
11 $ structural parameters and the incident wavelength, $\lambda $, which is
restricted to the $300-800$ nm range.

The shape of the asymmetric lens depicted in Fig. \ref{fig1}B is a polygon
consisting of the same number of vertices as in Fig. \ref{fig1}A. In this
case, however, we allow $22$ vertices to vary independently on both sides of
the vertical red line bisecting the NP. This results in $22$ structural
parameters and, as in the case of the symmetric lens, the incident
wavelength is determined as one additional independent parameter. The
incident pulse is of the form,
\begin{equation}
E_{\text{inc}}\left( t\right) =\sin ^{2}\left( \pi \frac{t}{\tau }\right)
\cos \left( 2\pi \frac{ct}{\lambda }\right) ,  \label{incident field}
\end{equation}%
where $\tau $ is the pulse duration, $c$ is the speed of light in vacuum,
and $\lambda $ is the incident wavelength. Throughout this subsection we
take $\tau =50$ fs with a total propagation time of $65$ fs, which ensures
that all plasmon polariton waves excited by the incident pulse decay within
a single FDTD run.

\begin{figure}[tbph]
\centering\includegraphics[width=\linewidth]{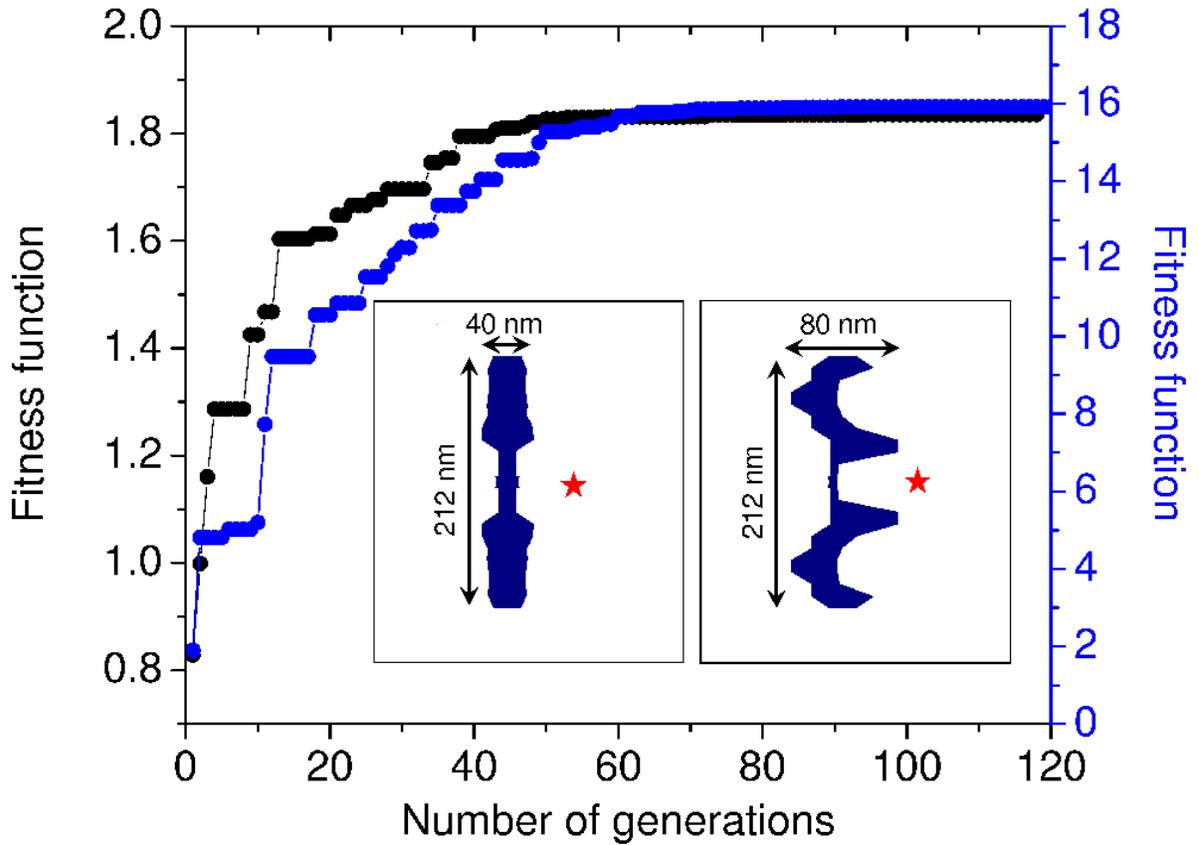}
\caption{(Color online) Two-dimensional silver nanolenses designed
by GA optimizations. The insets present the optimal silver NPs
(the left inset shows the case where symmetry is imposed whereas
the right inset shows the case where it is not). The main panel
illustrates the convergence of the GA iterations for the symmetric
(black circles with the values shown as left ordinate) and
asymmetric (blue circles with the values shown as right ordinate)
lenses. The fitness function in both cases is defined as the time
averaged EM energy (normalized with respect to the incident
energy) recorded at the focus (indicated by a red star). }
\label{fig2}
\end{figure}

Figure \ref{fig2} presents the optimization results. The main panel
illustrates the convergence of the GA iterations and the insets show the
optimal lenses. It is remarked that both the symmetric and the asymmetric
constructs tend to minimize the thickness of the central slab, suggesting,
at first sight, that a dimeric construct would be the best light localizer.
For EM enhancement near the surface of the metal construct, dimeric
configurations are expected, and were previously found,\cite{SchatzJCP2004}
to optimize the enhancement. In the case of a true lens, however, where the
focus is arbitrary and pre-specified, the dimeric construct is not the
optimal solution. In studies of the symmetric lens we found that the GA
converges to the structure shown in Fig. \ref{fig1}A whether or not the
thickness of the central slab is lower bounded. The thin wire joining the
top and bottom portions of the lens plays an essential role in the light
focussing dynamics. In simulations of the steady-state distributions of the
currents through the constructs, we found that the symmetric lens central
slab supports strong current flow in the vertical direction bouncing back
and forth between upper and lower parts of the lens. The importance of the
slab is also confirmed by independent simulations, in which we artificially
remove the slab from the lens. The enhancement factor in this case is
noticeably lower as compared to the GA optimized NP.

Interestingly, the optimal shapes obtained differ significantly for the
symmetric (left inset of Fig. \ref{fig2}) and asymmetric (right inset of
Fig. \ref{fig2}) cases. The two constructs differ also in their focussing
mechanism. The symmetric lens represents an analog of the dipole antenna
that was proven to efficiently enhance EM radiation at its center of
symmetry.\cite{CrozerAPL06} The asymmetric lens applies a more complex
approach to optimizing the enhancement and spatial localization of the
light, utilizing to that end, as discussed below, the long peaks marked in
the vicinity of the detection point and the two narrow summits at the top
and bottom of the lens.

\begin{figure}[tbph]
\centering\includegraphics[width=\linewidth]{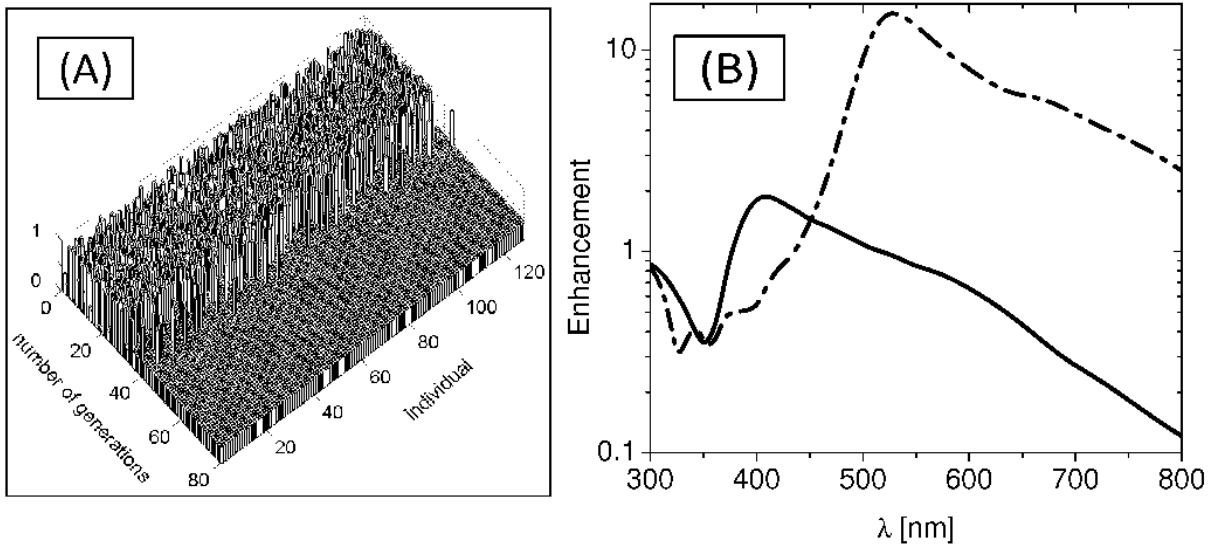}
\caption{Panel (A) shows the evolution of one of $11$ structural
parameters (in normalized units such that $P\in \left[ 0,1\right]
$) for the symmetric nanolens as a function of the number of
generations for all $128$ individuals used in the simulations.
Panel (B) gives the time averaged EM energy enhancement at the
focus (indicated in the insets of Fig. 2 as a red star) for the
optimal symmetric (solid curve) and asymmetric (dash-dotted curve)
lenses as a function of the incident wavelength.} \label{fig3}
\end{figure}

The use of a large number of parameters gives rise to rapid convergence of
the GA optimizations as illustrated in Fig. \ref{fig3}A, where we follow the
evolution of one of the structural parameters (normalized to range from 0 to
1) as the optimization progresses. At the start of the iterations (from
generation $1$ to about $25$) we observe large variations in the value of
the parameter among individuals of the same generation. As the GA converges,
however, the best solutions tend to have similar values for this particular
parameter. This means that individuals with this value are more likely to be
explored, resulting in an increase in the number of individuals with similar
values for this parameter as the GA progresses. At this point the major
structural features are defined, and only minor corrections are made to
reach an optimal solution. The optimal value of the parameter considered is
attained in less than $40$ generations. To verify that the algorithm
converges to the global extremum of the fitness function, we performed
calculations of the enhancement at the detection point for the optimal
symmetric and asymmetric lenses as a function of the incident wavelength, $%
\lambda $. The results of these simulations, presented in Fig. \ref{fig3}B,
show well defined extrema at the wavelengths to which the GA optimizations
converged.

\begin{figure}[tbph]
\centering\includegraphics[width=\linewidth]{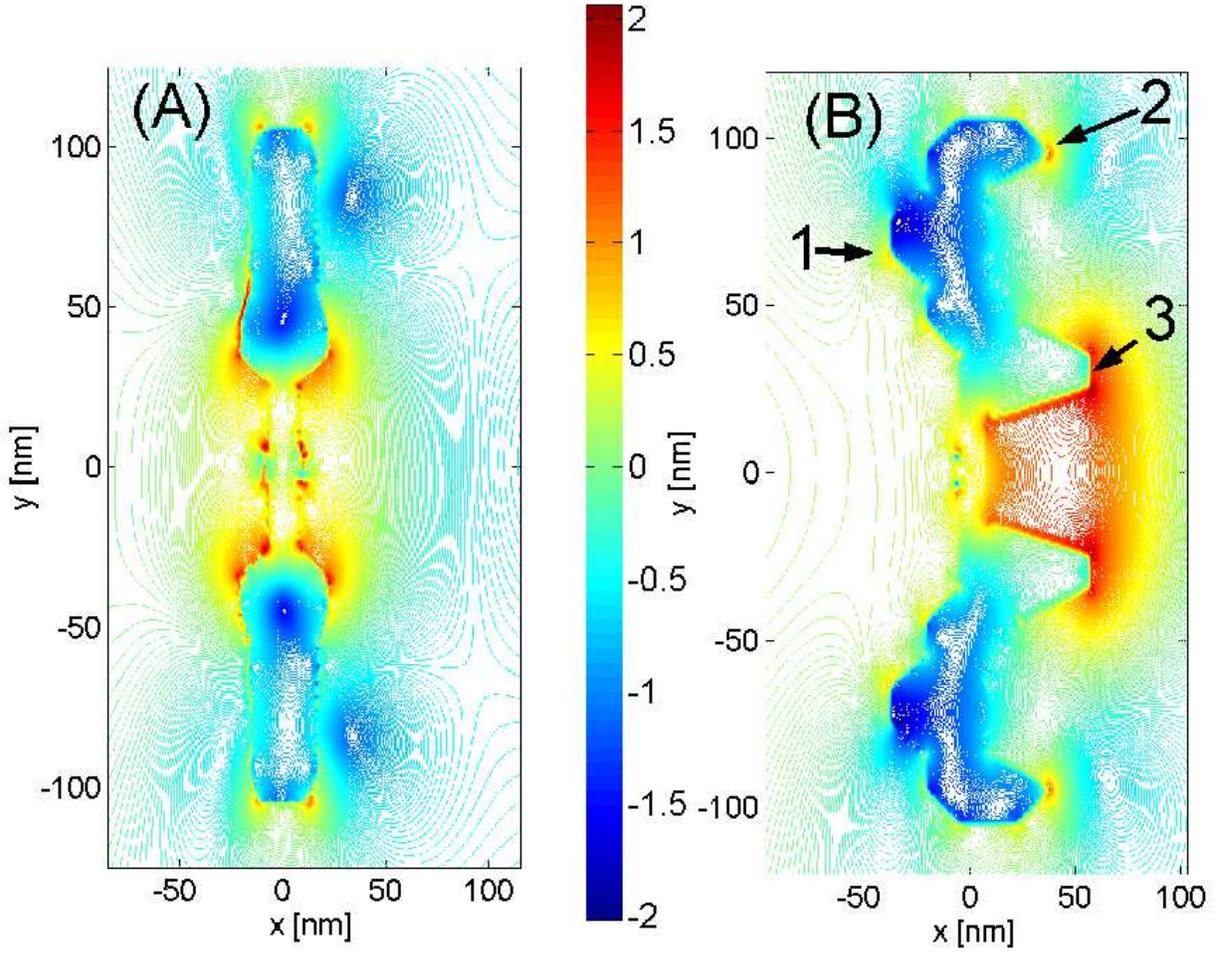}
\caption{(Color online) Spatial distributions of the steady-state
EM energy normalized with respect to the incident EM energy for
optimally designed symmetric (panel (A)) and asymmetric (panel
(B)) plasmonic lenses in logarithmic scale. Several geometric
features of the asymmetric lens are indicated by $1$, $2$, and $3$
and discussed in the text.} \label{fig4}
\end{figure}

The physics underlying the optimally designed plasmonic lenses can be
understood by calculation of the steady-state EM fields obtained via the
phasor function formalism (see, for instance, Chapter 8 of Ref. [%
\onlinecite{TafloveBook}]). The spatial distribution of the steady-state EM
energy shown in Fig. \ref{fig4}A for the symmetric lens confirms that the
optimal symmetric lens enhances the EM radiation at the detection point as a
nanoscale dipole antenna. A qualitative difference from the latter, however,
is the thin central slab, which, as discussed above plays an important role
in the focussing design. The asymmetric lens of Fig. \ref{fig4}B, by
contrast, applies a more complex strategy. Studies of the time evolution of
the EM intensity through the construct (see the upload-able animations\cite%
{movie} for illustration of the EM field focusing by the asymmetric lens)
illustrate that the incident radiation at the optimal frequency first
excites surface waves at the input (left) side of the lens and next, due to
the thin core, passes through the structure in the vicinity of its center.
The thin core strongly couples the plasmon waves on both sides of the lens.
By contrast to the dipole antenna, however, the asymmetric lens is concave
at the input side (feature $1$ in Fig. \ref{fig4}B), which serves to enhance
the EM radiation at the center. The incident field also excites localized
plasmon waves that travel around the lens and are accumulated at the sharp
corners marked by $2$. Finally, the two well pronounced peaks indicated as $3
$ in Fig. \ref{fig4}B act on the one hand as a cavity to eliminate loss of
the EM radiation passed through. On the other hand, they are strongly EM
coupled to the spikes marked $2$. We tested the importance of the feature $2$
by removing it from the optimal lens, finding, by recalculation of the
enhancement, a factor of $3$ lower enhancement at the focus.

\subsection{Optimal Birefringence}

\label{cross}

This subsection is motivated in part by recent studies,\cite{SukharevPRB07,
SukharevJPCC08} in which it was demonstrated that single L-shaped silver NPs
and arrays thereof exhibit strong birefringence, with important applications
in optics.\cite{NovotnyBook} As discussed in Sec. \ref{introduction},
polarization effects requires 3D simulations. The combination of the GA with
FDTD in full 3D, however, present a numerical challenge since 3D FDTD
calculations are numerically costly. In this subsection we follow the
optimization approach developed for nanoplasmonics in Ref. [%
\onlinecite{SukharevJPB07}]. Here an individual is evaluated by partitioning
the FDTD grid onto $256$ processors and implementing point-to-point
communication MPI subroutines as discussed in Ref. [%
\onlinecite{SukharevPRB07}]. Average execution time of GA iterations for $150
$ generations is nearly $5$ days.

We consider a periodic array of $X$-shaped silver NPs, as depicted
schematically in Fig. \ref{fig1}, which is irradiated by an $x$-polarized
incident plane wave propagating along the $z$-axis. The target of the
optimization is to enhance the birefringence of the periodic array, i.e., to
maximize the ratio of the time averaged EM energy in the field component
perpendicular to the incident field polarization, $W_{y}=E_{y}^{2}$, to the
corresponding energy in the field component parallel to the incidence
polarization, $W_{x}=E_{x}^{2}$. As schematically depicted in Fig. \ref{fig1}%
, the free parameters are the lengths of the four arms of the NP, $%
P_{1},\ldots P_{4}$, the angle of rotation, $P_{5}$, and the incident
wavelength, $\lambda $. The number of individuals in each generation is $8$
and the incident pulse duration is $\tau =20$ fs, with a total propagation
of $23$ fs. The height and the thickness of all arms of the NP are $30$ nm.
The ratio $W_{y}/W_{x}$ is averaged over an $xy$-plane parallel to the array
plane on the output side at a distance of $656$ nm from the array. The limit
of $W_{y}/W_{x}\rightarrow 0$ corresponds to poor or no birefringence,
whereas $W_{y}/W_{x}\rightarrow 1$ corresponds to equal magnitudes of the
incident polarization and the field component that is purely induced by the
array.

\begin{figure}[tbph]
\centering\includegraphics[width=\linewidth]{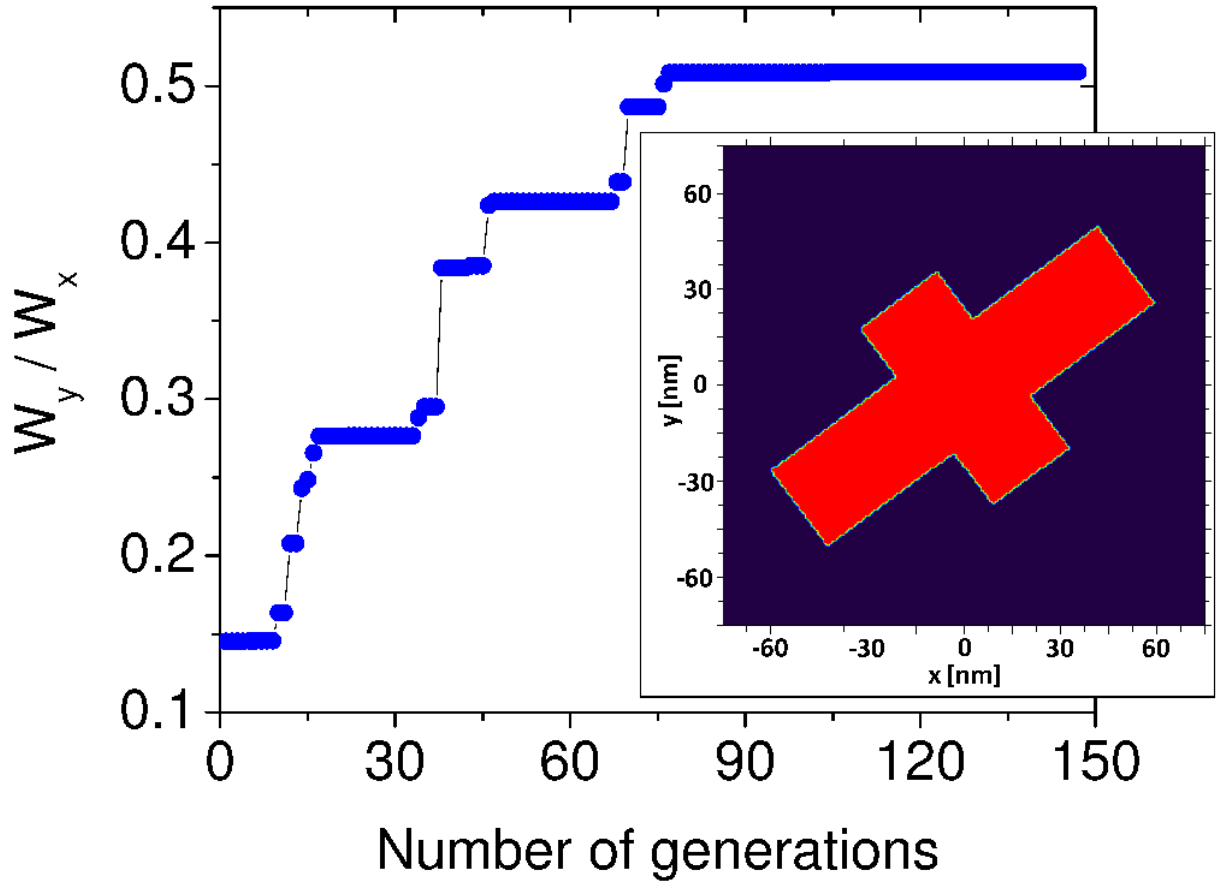}
\caption{(Color online) GA design of a three-dimensional plasmonic
array to depolarize light in a predetermined fashion. The main
panel shows the evolution of the fitness function, $W_{y}/W_{x}$,
with the number of generations. The inset depicts the optimal unit
cell in the $xy$-plane bisecting the array, where spatial regions
occupied by silver are shown in red and free space regions are
shown in dark blue.} \label{fig5}
\end{figure}

The main panel of Fig. \ref{fig5} presents the evolution of the ratio, $%
W_{y}/W_{x}$, with the number of GA generations and the inset shows the
optimized unit cell of the array after $149$ generations. We have that for
the incident pulse duration considered, the optimized ratio exceeds $0.5$,
corresponding to strong depolarization of the EM field by the array. Since
the enhanced birefringence relies on resonant excitation of localized
plasmon waves, increase of the incident pulse duration at the optimal
wavelength is expected to further increase the $W_{y}/W_{x}$ ratio. We
confirmed this expectation by calculating the birefringence ratio as a
function of the incident wavelength for the geometry to which the GA
converged and found that $W_{y}/W_{x}\approx 0.62$ for laser pulses with
durations $100$ fs or longer. We remark that the optimal incident wavelength
(the wavelength to which the algorithm converged), $\lambda =554.3$ nm,
corresponds to the global maximum of $W_{y}/W_{x}$, as in the case of two
dimensional simulations (see Fig. \ref{fig4}B and the corresponding
discussion).

\begin{figure}[tbph]
\centering\includegraphics[width=\linewidth]{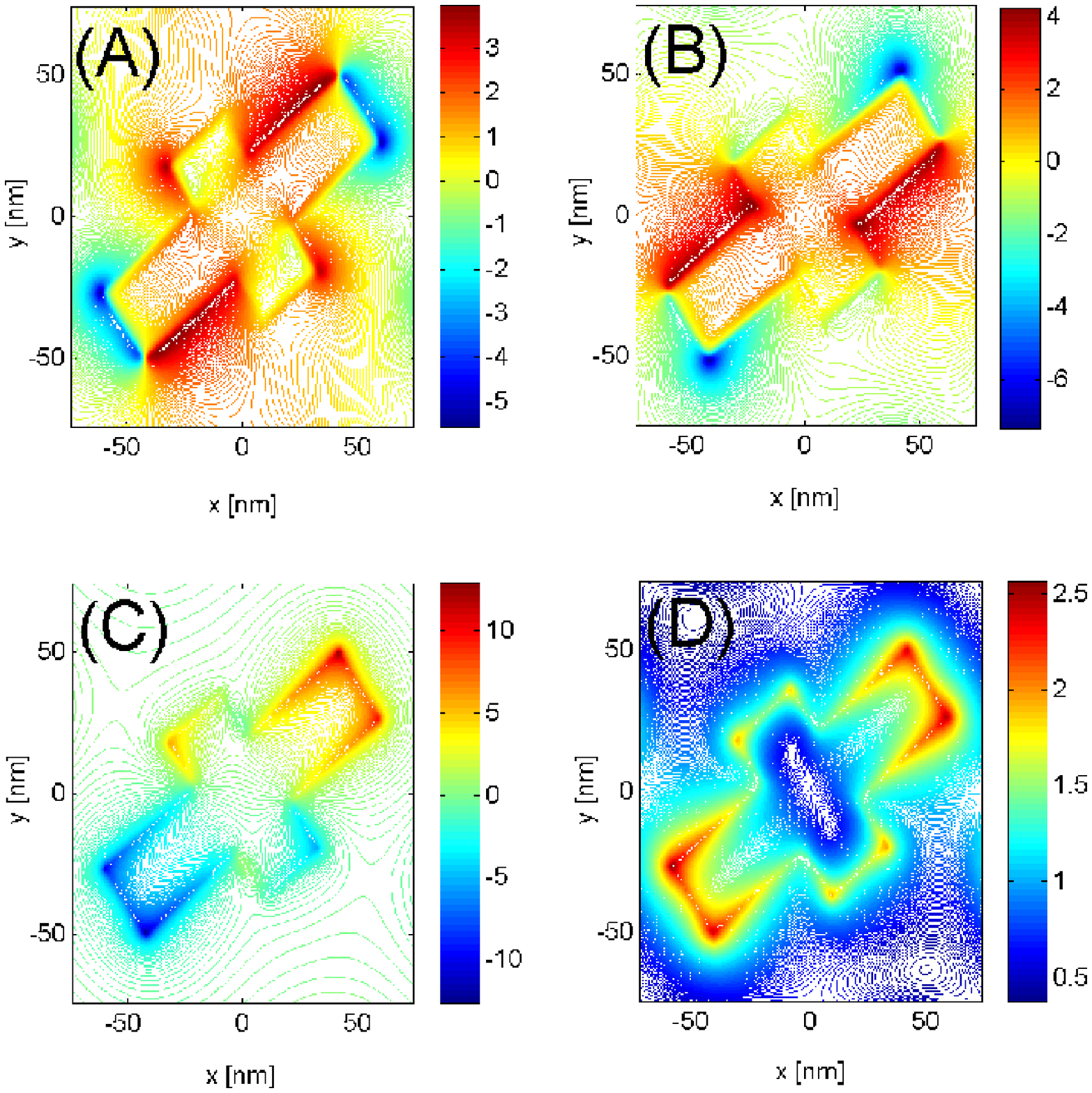}
\caption{(Color online) Study of the optimized construct resulting
from the design of Fig. 5. Spatial distributions of the
steady-state electric field components $E_{x}$ (A), $E_{y}$ (B)
and $E_{z}$ (C) (normalized to the incident field amplitude) as
functions of $x$ and $y$ on the output side the
array at a distance of $5$ nm. Panel D shows the EM intensity distribution, $%
E_{x}^{2}+E_{y}^{2}+E_{z}^{2}$, on a logarithmic scale (normalized to the
incident peak intensity).}
\label{fig6}
\end{figure}

The optimally designed array gives rise to complex EM field dynamics, as
shown in Fig. \ref{fig6}, where we plot the steady-state solutions of the
Maxwell equations on the output side of the array at a distance of $5$ nm.
It is clearly seen that the induced polarization component depends markedly
on the distance from the array and is particularly strong at short
distances. The induced $E_{z}$ component ($z$ with respect to the reference
frame of the incident $k$-vector) shown in Fig. \ref{fig6}C, corresponds to
redirection of the scattered field by the array and rapidly decays with
distance from it.

Simulations of the EM eigenmodes of the array (not shown) reveal the origin
of the strong birefringence found. At the resonant wavelength, the total
field is determined by the field induced collective oscillations of the
electrons confined in the particles, which in turn is largely determined by
the polarizability tensor of the particle, hence its shape and orientation
with respect to the direction of incidence. An interesting observation is
the large relative phase, $\delta \varphi _{xy}$, between $E_{x}$ and $E_{y}$
in the far field region. The phase along with the ratio $W_{y}/W_{x}$ fully
defines the polarization of the scattered EM field. For the optimal
parameters the phase, $\delta \varphi _{xy}$, approaches $211^{0}$ at a
distance of $656$ nm on the output side of the array and is almost
independent of $x$ and $y$.

It is instructive to re-express the polarization of the scattered field in
terms of Stokes parameters, defined as\cite{SharmaOpticsBook2006}%
\begin{equation}
\hat{S}=\left(
\begin{array}{c}
S_{0} \\
S_{1} \\
S_{2} \\
S_{3}%
\end{array}%
\right) =\frac{1}{E_{x}^{2}+E_{y}^{2}}\left(
\begin{array}{c}
E_{x}^{2}+E_{y}^{2} \\
E_{x}^{2}-E_{y}^{2} \\
2E_{x}E_{y}\cos \delta \varphi _{xy} \\
-2E_{x}E_{y}\sin \delta \varphi _{xy}%
\end{array}%
\right) ,  \label{Stokes}
\end{equation}%
where $E_{x}$ and $E_{y}$ are the steady-state solutions of the Maxwell
equations for the optimal array. Whereas the incoming incident field is
described by the Stokes parameters $\left( S_{0}^{\left( \text{in}\right)
},S_{1}^{\left( \text{in}\right) },S_{2}^{\left( \text{in}\right)
},S_{3}^{\left( \text{in}\right) }\right) =(1,1,0,0)$, the strong
depolarization induced by the array results in $\left( S_{0}^{\left( \text{%
out}\right) },S_{1}^{\left( \text{out}\right) },S_{2}^{\left( \text{out}%
\right) },S_{3}^{\left( \text{out}\right) }\right) =(1,0.24,-0.83,-0.50)$,
corresponding to left elliptical polarization. We remark the symmetry
breaking between the left and right polarized components of the incident
(linearly-polarized) light that is induced by the chirality of the system.

\section{Conclusion}

\label{conclusions}Our goal in the work discussed in the previous sections
has been to extend and apply a recently developed\cite%
{SukharevNanolett06,SukharevPRB07} approach for optimal design of plasmonic
nanoconstructs, both as a route to new insights into plasmonic physics and
as a practical tool for fabrication of optical nanodevices with desired
functionalities. To that end we introduced multi-parameter GA optimizations
in both two and three dimensions.

Using fast FDTD algorithms, we first applied the GA to the problem of
focusing of EM radiation by subwavelength scaled silver lenses. We found
that symmetry plays a major role in light focusing and directing in the
nanoscale. The detailed mechanisms by which a lens constrained to be
symmetric and one allowed to assume asymmetry guide the EM field in space
were explored and the role of coherence was noted. GA optimizations were
next applied to three-dimensional periodic plasmonic systems to manipulate
the polarization of the EM field. The ability of properly structured but
rather simple cross-shaped silver NPs to convert a linearly polarized field
into left or right elliptically polarized field with comparable x- and
y-components was illustrated.

A variety of potential extensions and applications of the numerical tool
developed above can be envisioned. One of the goals of our current research
is to develop nanoscale light sources with desired phase and polarization
properties that would allow the extension of coherent control to the
nanodomain. Another is the numerical design of functional plasmonic
nanodevices where the structural parameters of the material system have been
optimized to yield a specific function. These include sensors, enhanced
solar cells, and efficient constructs for plasmonic enhancement of
spectroscopic signals. A third topic of ongoing work is the application of
tools of phase, polarization and optimal control to introduce new function
into nanoplasmonics, including broken symmetry elements and ultrafast
elements.

\begin{acknowledgments}
This research was supported in part by the NCLT program of the National
Science Foundation ESI-0426328 at the Materials Research Institute of
Northwestern University. The numerical work used the resources of the
Argonne Leadership Computing Facility at Argonne National Laboratory, which
is supported by the Office of Science of the U.S. Department of Energy under
Contract No. DE-AC02-06CH11357.
\end{acknowledgments}

\newpage

\newpage

\newpage

\newpage

\newpage

\newpage


\begin{thebibliography}{99}
\bibitem{reviews} (1) S. A. Maier, M. L. Brongersma, P. G. Kik, S. Meltzer,
A. A. G. Requicha, and H. A. Atwater, Adv. Mater. (Weinheim, Ger.) \textbf{13%
}, 1501 (2001); (2) S. Link and M. A. El-Sayed, Annu. Rev. Phys. Chem.
\textbf{54} 331 (2003); (3) E. Hutter and J. H. Fendler, Adv. Mater.
(Weinheim, Ger.) \textbf{16}, 1685 (2004); (4) E. Ozbay, Science \textbf{113}
189 (2006).

\bibitem{MaierBook07} S. A. Maier, \textit{Plasmonics: Fundamentals and
Applications}, Chapter 8 and references therein, (Springer, New York, 2007).

\bibitem{AtwaterReview05} S. A. Maier and H. A. Atwater, J. Appl. Phys.
\textbf{98}, 011101 (2005).

\bibitem{VanDuyneScience04} R. P. Van Duyne, Science \textbf{306}, 985
(2004).

\bibitem{HlaReview05} S.-W. Hla, J. Vac. Sci. Technol. B \textbf{23}, 1351
(2005).

\bibitem{VanDuyneReview04} A. J. Haes and R. P. Van Duyne, Expert Rev. Mol.
Diagn. \textbf{4}, 527 (2004).

\bibitem{BarnesReview07} W. A. Murray and W. L. Barnes, Adv. Mater.
(Weinheim, Ger.) \textbf{19}, 3771 (2007).

\bibitem{VanDuyneJPCB2001} C. L. Haynes and R. P. Van Duyne, J. Phys. Chem.
B \textbf{105}, 5599 (2001).

\bibitem{SukharevNanolett06} M. Sukharev and T. Seideman, Nanolett. \textbf{6%
}, 715 (2006).

\bibitem{SukharevJCP06} M. Sukharev and T. Seideman, J. Chem. Phys. \textbf{%
124}, 144707 (2006).

\bibitem{SukharevJPB07} M. Sukharev and T. Seideman, J. Phys. B \textbf{40},
S283 (2007).

\bibitem{TafloveBook} A. Taflove and S. C. Hagness, \textit{Computational
Electrodynamics:\ The Finite-Difference Time-Domain Method}, 3rd ed. (Artech
House, Boston, 2005).

\bibitem{GreyPRB03} S. K. Gray and T. Kupka, Phys. Rev. B \textbf{68},
045415 (2003).

\bibitem{Ziolkowski95} J. B. Judkins and R. W. Ziolkowski, J. Opt. Soc. Am.
A \textbf{12}, 1974 (1995).

\bibitem{AgDrude} Parameters in Eq. (\ref{Drude model}) for silver used in
our simulations are: $\varepsilon _{\infty }=8.926$, $\omega
_{p}=1.7601\times 10^{16}$ rad/sec, and $\Gamma =3.0841\times 10^{14}$
rad/sec.

\bibitem{Berenger} J.-P. Berenger, J. Comput. Phys. \textbf{114}, 185 (1994).

\bibitem{SadikuBook} M. N. O. Sadiku, \textit{Numerical Techniques in
Electromagnetics}, 2nd ed. (CRC, Boca Raton, FL, 2001).

\bibitem{HauptGAbook} R. L. Haupt and S. E. Haupt, \textit{Practical Genetic
Algorithms} (John Wiley \& Sons, 2004).

\bibitem{SchatzJCP2004} E. Hao and G. C. Schatz, J. Chem. Phys. \textbf{120}%
, 357 (2004).

\bibitem{CrozerAPL06} E. Cubukcu, E. A. Kort, K. B. Crozier, and F. Capasso,
Appl. Phys. Lett., \textbf{89}, 093120 (2006).

\bibitem{movie} http://wealth.qserty.ru/Physics/Nano/asymmetric\_lens.zip
This movie shows scattering of EM plane wave by the optimally designed
asymmetric lens (see Fig. 2 and discussion in the text). EM plane wave
propagates from left to the right and is polarized along vertical axis. Each
frame of the animation presents normalized EM energy in logarithmic scale as
a function of spatial coordinates.

\bibitem{SukharevPRB07} M. Sukharev, J. Sung, K. G. Spears, and T. Seideman,
Phys. Rev. B \textbf{76}, 184302 (2007).

\bibitem{SukharevJPCC08} J. Sung, M. Sukharev, E. M. Hicks, R. P. Van Duyne,
T, Seideman, and K. G. Spears, J. Phys. Chem. C (in press, 2008).

\bibitem{NovotnyBook} L. Novotny and B. Hecht, \textit{Principles of
Nano-Optics} (Cambridge University Press, Cambridge, England, 2006).

\bibitem{SukharevJCP07} M. Sukharev and T. Seideman, J. Chem. Phys. \textbf{%
126}, 204702 (2007).

\bibitem{SharmaOpticsBook2006} K. K. Sharm, \textit{Optics: principles and
applications} (Elsevier, 2006).
\end{thebibliography}
\end{document}